\documentclass[preprint]{aastex}
\input{psfig}

\newcommand{\nh}{$N_{\rm H~}$}
\newcommand{\sax}{{\it BeppoSAX~}}

\newcommand{\comptel}{{Comptel~}}

\newcommand{\cena}{Centaurus A~} 
\begin{document}

\title{BeppoSAX Observations of \cena: the Hard Continuum and 
the Iron Line Feature}

\author{P. Grandi}
\affil{Istituto di Astrofisica Spaziale e Fisica Cosmica, CNR,
Sezione di Bologna, Via Gobetti 101, I-40129 Bologna, Italy}
\author{M. Fiocchi}
\affil{Asi Scientific Data Center, Frascati, Italy}
\author{C. G. Perola}
\affil{Universita' degli Studi ``Roma Tre'', Roma, Italy}
\author{C. M. Urry}
\affil{Department of Astronomy, Yale University, New Haven, USA}
\author{L. Maraschi}
\affil{Osservatorio Astronomico di Brera, Milano, Italy}
\author{E. Massaro}
\affil{Universita' degli Studi ``La Sapienza'', Roma, Italy}
\author{G. Matt}
\affil{Universita' degli Studi ``Roma Tre'', Roma, Italy}
\author{A. Preite-Martinez}
\affil{Istituto di Astrofisica Spaziale e Fisica Cosmica, CNR,
Roma, Italy}
\author{H. Steinle and W. Collmar}
\affil{Max Plank Institute f\"ur extraterrestrische Physik, 
Garching, Germany}

\begin{abstract}\small
The radio galaxy \cena was observed by the BeppoSAX 
satellite five times from 1997 to 2000. From 
July 6 1999 to August 17 1999, the source was also simultaneously pointed
by  COMPTEL on-board of the $\gamma-ray$ satellite CGRO. 
\cena has a complex spectrum with multiple extended components 
and a strongly absorbed (\nh$\sim 10^{23}$ cm$^{-2}$) 
nucleus well fitted by a power law ($\Gamma \sim 1.8$) which bends
at high energies. When the \sax and COMPTEL observations are combined
together, an exponential cutoff with e-folding energy $\sim 1000$  keV 
gives an adequate description of the spectral steepening.
A complex feature in emission at 6-7 keV is resolved into two 
Fe $K_{\alpha}$ components, one narrow cold line and 
an ionized line centred at 6.8 keV .
Significant variations have been observed in the iron feature,
with the less prominent ionized line seemingly being the only one 
responsible for them: its variations do not appear to correlate with the strength
of the continuum. 
The high energy cutoff and the Fe feature suggest 
the presence of an accretion flow in the \cena nucleus. 
However the absence of a significant reflection, the narrowness 
of the cold line as well as the  
lack of correlation between the continuum and 
6.8 keV  line variations disfavour a standard cold/ionized 
thin disk (at least in the inner regions). 
A more plausible configuration might be 
a hot thick optically thin accretion flow surrounded by material 
with different opacities.
Finally, we note that high energy break observed by \sax and COMPTEL 
could be also reasonably explained by  Inverse Compton radiation from a jet.
If this is the case, a structured jet with outer slow layers surrounding 
a beamed inner region is necessary  to explain the strong Fe feature 
observed by \sax.

\end{abstract}
\normalsize

\newpage
\section{INTRODUCTION}

Thanks to its proximity (D=3.4 Mpc, Israel 1998) \cena is a well-known 
radio-loud AGN extensively study from the radio to $\gamma$-ray band.
At optical wavelengths, it coincides with the giant elliptical 
galaxy NGC5128, bisected by a prominent dust line. 
At radio frequencies, it shows a FR I morphology
with several emitting features: two giant (250 kpc) outer lobes, 
a northern middle lobe extending to 
30 kpc, two inner lobes and central jets (of 1-5 kpc linear size) and a core 
(0.01 pc) with associated nuclear jet extending to about 1 pc (Israel 1998 and references therein).

Before the launch of the X-ray satellite Chandra, the X-ray counterparts 
of the radio nucleus, of the jet and the northern radio lobe have already 
been revealed, as well as a faint diffuse emission within 6$^\prime$ 
from the nucleus, probably from hot interstellar medium, and two ridges 
of thermal emission along each edge of the dust lane, probably associated with 
Population I binary systems in the disk (Feigelson et al. 1981,  
Morini et al. 1989, D\"{o}bereiner et al. 1996, Turner et al. 1997). 
With Chandra the X-ray spatial structure has been studied in greater detail.
This satellite has shown a nuclear extension of about 5 pc 
(a few tenths of an arcsecond), an X-ray counter-jet, weak emission 
from the southern radio lobe and has also detected 63 point-like galactic 
sources (Kraft et al. 2001a, Kraft et al. 2001b). In particular the jet has 
been resolved in 
a diffuse emission extending continuously from 60 pc to 4 
kpc and in 31 discrete compact knots, well fitted with absorbed 
power laws ($\Gamma\sim 2.0-2.5$) (Kraft et al. 2002).
Steinle et al. (2000) also reported the discovery of a bright X-ray transient 
at about 2.5 arcminutes from the nucleus. This source could be an X-ray binary 
in NGC5128 with a luminosity $3 \times 10^{39}$ 
erg sec$^{-1}$ in the 0.1-2.4 keV band. 

Because of the large extinction suffered by the nucleus, 
the extended components dominate the emission at soft X-ray
energies. Above 3~keV the nuclear source becomes dominant.
The nuclear  spectrum between 3-10~keV has been extensively studied 
in the past by all the main X-ray satellites. It is generally modeled with a 
power law heavily absorbed by dense and probably stratified matter 
(\nh $\sim 10^{-23}$ cm$^{-2}$) and by a fluorescence iron line (Mushotzky 
et al. 1978, 
Wang et al. 1986, Morini et al. 1989, Miyazaki et al 1996, Turner et al. 1997,
Sugizaki et al. 1997). In contrast with Seyfert galaxies, the iron line of \cena
is not associated with a strong Compton reflection (Miyazaki et al. 1996, 
Wozniak et al. 1998, Rotschild et al. 1999), as one would expect if the
line were produced by cold and optically thick gas. This convinced many authors
that the Fe feature is produced by the cold dense, but optically thin gas,
which is also responsible for the observed photoelectric absorption (\nh)
surrounding the X-ray source  (Miyazaki et al. 1996, Wozniak et al. 1998, 
Miyazaki et al 1996, Benlloch et al. 2001). 

Hard X-ray and soft $\gamma$-ray observations performed with the GRO satellite
showed a steepening of the power law at high energies.
Adopting a power law with an exponential cut-off, the e-folding energy ranged from 300 to 700 keV 
depending on the brightness state of the source. The high energy spectrum 
hardened with decreasing intensity (Kinzer et al. 1995).
Combining the data from all the GRO instruments, OSSE COMPTEL and EGRET, 
Steinle et al. (1998) showed that a further steepening of the spectrum occurs
between the MeV (COMPTEL) and  GeV (EGRET) energies.

Here we present the results of 5 \sax observations of \cena, two of them
simultaneously performed with the COMPTEL instrument on-board the CGRO 
satellite. Preliminary results concerning the first two pointings were 
presented by Grandi et al. (1998).

\section{Observations and Data Analysis}

\sax observed \cena on 5 occasions from 1997 to 2000 (Table 1) 
and on two occasions operated simultaneously with the COMPTEL 
instrument on board of CGRO (Table 2).

LECS, MECS and PDS data were reduced  following the standard procedures 
described by Fiore Guainazzi and Grandi (1999; F99).
Both the LECS and the MECS spectra were accumulated in circular regions 
of 4' radius. Because of a failure in the MECS unit 1 
in 1997 May 6, data from all three MECS units were available only for 
the first observation.
Background spectra were extracted from blank fields observations at the
same position of the source in the detectors.
Since the source is strongly dominated by the background below 0.4 keV, 
and the response matrix is less reliable above 4 keV, the
LECS energy range was restricted to 0.4-4 keV.
The PDS net source spectra and light curves were
produced by subtracting the off- from the on-source products. 
The PDS spectra were obtained 
with the background rejection method based on fixed rise time thresholds.
Due to the brightness of the source, we preferred to use the standard PDS 
reduction procedure to avoid systematic errors in the spectral slopes
(F99). Publicly available matrices were used for all the instruments.
We binned the LECS, MECS and PDS spectral files proportionally to 
the instrumental resolution (F99).

The COMPTEL data have been obtained by merging the short observation
periods 821, 822, 823, 824, 825, 826 (one week each) into one spectrum.  
All the COMPTEL observations have been made in a
special "solar mode". However,  this was taken into account in the 
data reduction and had no influence on the resulting COMPTEL spectral data. 
The derived intensities (fluxes)
are comparable to the values measured with COMPTEL in viewing periods
before and after the measurements discussed here.

\begin{deluxetable}{llcccccc}
\tabletypesize{\small}
\tablecolumns{8}
\tablecaption{BeppoSAX Observations$^a$}
\tablehead{
\multicolumn{2}{c}{\bf Date}
&\multicolumn{3}{c}{\bf Exposure time}
&\multicolumn{3}{c}{\bf Count sec$^{-1}$}\\
&
&\multicolumn{1}{c}{LECS}
&\multicolumn{1}{c}{MECS}
&\multicolumn{1}{c}{PDS}
&\multicolumn{1}{c}{LECS}
&\multicolumn{1}{c}{MECS}
&\multicolumn{1}{c}{PDS}\\
year-month-day 
&\colhead{MJD} 
&\colhead{[ksec]}
&\colhead{[ksec]}
&\colhead{[ksec]}
&\colhead{[0.4-4 keV]}
& \colhead{[1.5-10 keV]}
&\colhead{[15-200 keV]}
}
\startdata
1997-02-20    & 50499 &17 &34 &15 &0.201$\pm0.003$&1.660$\pm0.007$&5.20$\pm0.07$\\
1998-01-06 & 50819 &21 &54 &23 &0.242$\pm0.004$&2.204$\pm0.006$&6.93$\pm0.05$ \\
1999-07-10 & 51369 &14 &39 &18 &0.236$\pm0.004$&2.053$\pm0.007$&6.35$\pm0.06$\\
1999-08-02 & 51392 &11 &36 &16 & 0.192$\pm0.004$&1.680$\pm0.007$&5.24$\pm0.06$\\
2000-01-08& 51551 &12 &34 &18 & 0.224$\pm0.004$&2.008$\pm0.008$&6.39$\pm0.06$\\

\enddata
\tablenotetext{a}{The count rates are background subtracted,
and their uncertainties are 1-$\sigma$.\\
MECS count rates refer to MECS2+MECS3 units}

\label{t:t1}
\end{deluxetable}

\begin{deluxetable}{llcccccc}
\tabletypesize{\small}
\tablecolumns{6}
\tablecaption{COMPTEL Observation}
\tablehead{
\multicolumn{1}{c}{\bf Start Time}
&\multicolumn{1}{c}{\bf End Time}
&\multicolumn{4}{c}{\bf Log [Flux (Jy Hz)]}\\
&&&&\\
year-month-day & year-month-day 
&\multicolumn{1}{c}{ [0.75-1 MeV]}
&\multicolumn{1}{c}{[1-3 MeV]}
&\multicolumn{1}{c}{[3-10 MeV]}
&\multicolumn{1}{c}{[10-30 MeV]}\\

}
\startdata
1999-07-06 & 1999-08-17 
&$<13.97$
&13.44$^{+0.12}_{-0.16}$
&$<13.23$ 
&$<13.34$
\enddata
\label{t:comptel}
\end{deluxetable}
\normalsize
We used {\sc xspec} 11 to fit the \sax data. 
The parameter uncertainties correspond to $90\%$  confidence for 
two interesting parameters, i.e., $\Delta \chi^2=+4.61$. 
LECS, MECS and PDS spectra were simultaneously fitted,
with the PDS to MECS ratio allowed to vary between 0.83-0.89,
while the LECS to MECs ratio was left free, on accounts
of possible flux variations and the fact that the 
LECS and MECS data only partially overlap in time. 
The significance level of a model was obtained from the $F$-test:
we assumed that a F probability larger 99.5$\%$  implied 
a significant improvement of the fit.
The error bars for the count rates in the figures are 1-$\sigma$.

\section{\sax Temporal Analysis }

\cena was in a relatively low state of brightness during our observations.
The observed fluxes, $F_{\rm 3-12~keV}\sim 0.2-0.3\times10^{-9}$ erg 
cm$^{-2}$ sec$^{-1}$, are within the range observed in the last 10 years
and lower by a factor $\sim 5$ than during the outburst in 1974-75 (Turner et al. 1997).

For each observation, time variability in bins of 1000, 5000 and 10000 s was 
systematically searched for, using the LECS (0.4-4 keV), 
MECS (1.5-10 keV) and PDS (15-250 keV) light curves.
A standard $\chi^2$ test was
applied to the average count rate in each energy band.
We accepted the source as variable if the probability of being constant was less
than $10^{-3}$. Fast variability was detected in the 1998 MECS observation,
when a drop by $\sim 10\%$ took place in about 8 hours, as shown in Figure 1 ({\it upper panel}).
The two MECS spectra, extracted before and after this event, do not display
any significant change of form, as demonstrated by their ratio, shown
in Figure 1 ({\it lower panel}).

When comparing the mean count rate of the five observations (Figure 2),
modest flux differences of about 10-25$\%$ are present in both the MECS and the PDS, 
with a $\chi^2$ probability less than $10^{-3}$ for the null hypothesis in each of the 
1.5-4 keV, 4-10 keV and 15-250 keV light curves. 
Below $<1.5$ keV no significant differences were observed, 
($\chi^2_\nu=1.27$, P$_{\chi^2}=0.28$), as expected, 
being the soft photons mainly produced in extended regions.

\begin{figure}
\begin{center}
\vbox{
\psfig{figure=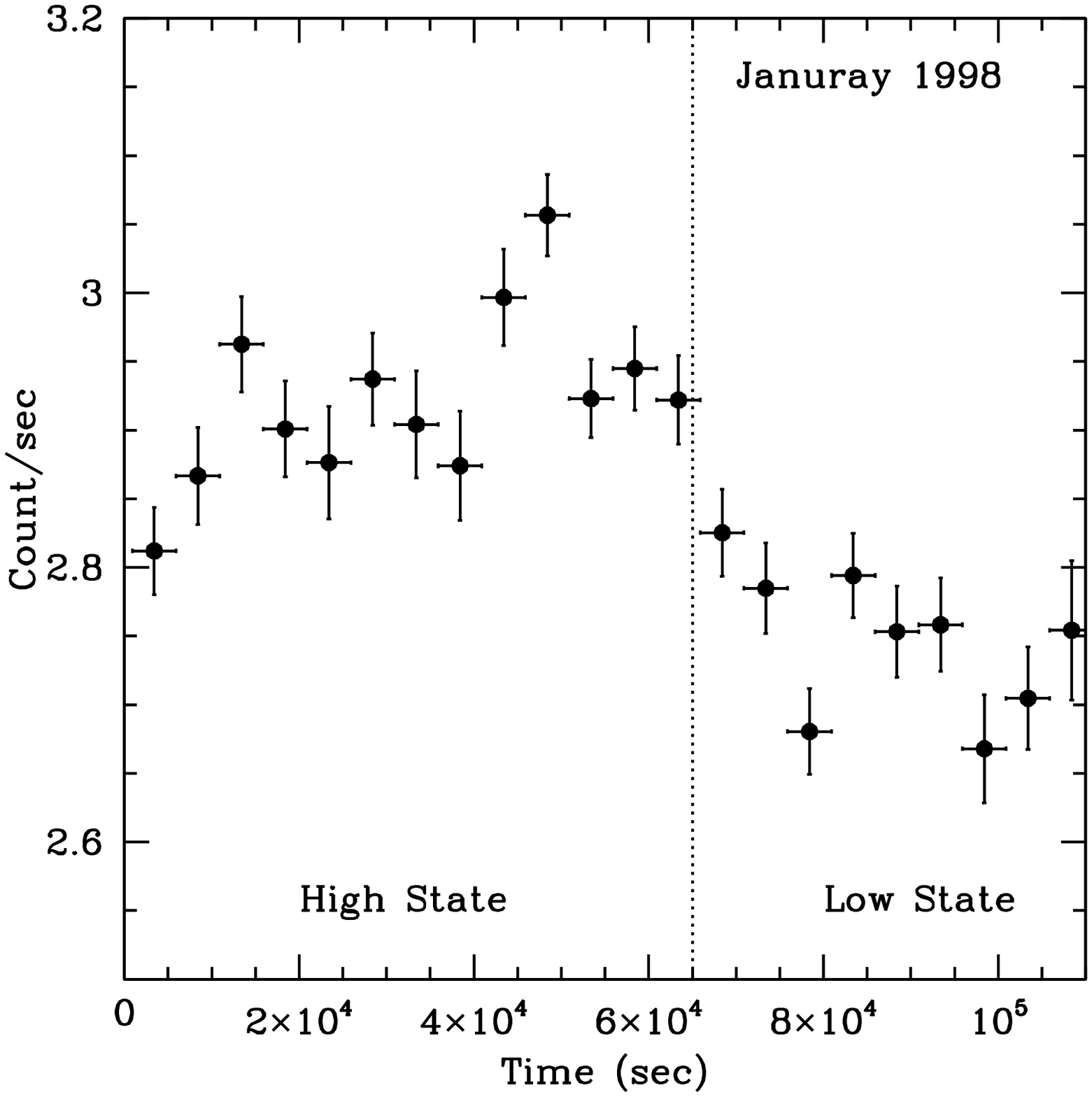,height=8.0cm,width=8.0cm}
\psfig{figure=figure1b.ps,height=7.5cm,width=8.0cm,angle=-90}
}
\vspace{-0.1cm}
\caption{\footnotesize{({\it Upper Panel})-- Fast flux 
variability observed on 8 January 1998 (temporal bins of 5000 sec) 
between 1.5 and 10 keV . 
({\it Lower Panel}) -- Ratio between the MECS spectra of 1998 
collected in the high and low states. The ratio is constant across the 1.5-10~keV band.} 
}
\end{center}
\end{figure}

\begin{figure}
\begin{center}
\vbox{
\psfig{figure=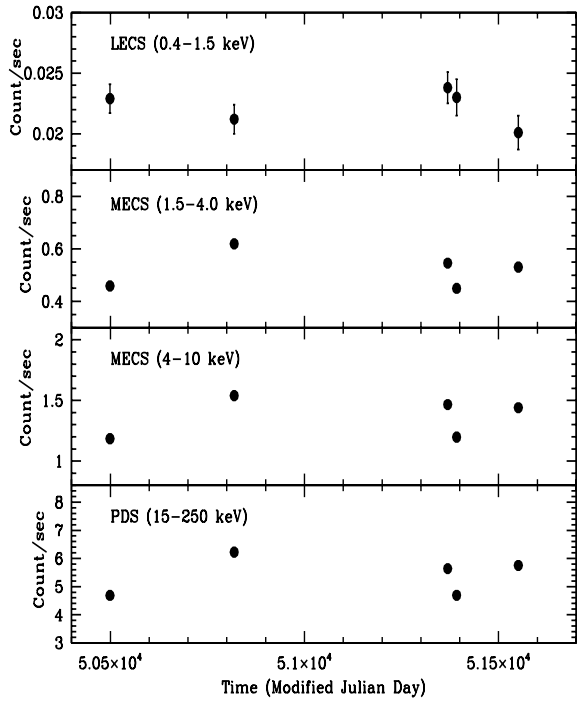,height=10.0cm,width=8.0cm}}
\vspace{-0.1cm}
\caption{\footnotesize{Long term \sax light curves collected 
from 1997 to 2000. Small amplitude flux variations 
characterize the medium and hard energy band. The error bars in 
the MECS and PDS light curves are smaller than the symbol size.}}
\end{center}
\end{figure}

\section{Spectral Analysis: The average Spectrum}

\cena is a source known for its spectral complexity. 
Several extranuclear components are required to fit the soft spectrum,
which \sax could not resolve. Since our goal is to exploit these data to
investigate the hard nuclear component, in our spectral analysis 
we followed previous investigations to model their contributions.
We further combined to start all five observations, in order to
achieve the highest signal to noise ratio. This procedure can,
in principle, be risky, since the source can change its
spectrum from one observation to another. However in our case, 
the nuclear continuum (above 4 keV) 
did not significantly vary in shape from one observation to another 
(as shown a posteriori,
but also judged a priori from spectral ratios), allowing us 
to take advantage of the high quality of the average spectrum up to 250 keV.

\subsection{\it Extended Emission and Nuclear Continuum}

The average LECS, MECS and PDS spectra have been obtained by combining 
all the five observations of \sax. 
Merging spectra of a source (in particular if extended) needs some precautions.
The data of the MECS unit 1, available only in 1997, were not included.
To avoid problems due to the different point spread 
functions\footnote{At 1.5 keV 
the 95$\%$ of LECS and MECS count/sec of a point-like source are contained 
within a region of 5.5 and 3.5 arcminutes, respectively.},
the MECS data below 3 keV were not considered and the soft extended 
components were modeled with the LECS alone (0.4-3.0 keV).

The study of the nuclear continuum was initially performed 
excluding the iron line region (from 5.5 to 7.5 keV).
Due to our inability to perform a spatially resolved analysis of the extended
emission, we followed the study  of Turner et al. (1997) based on 
ROSAT and ASCA observations.
They showed that the soft emission can be fitted 
with two thermal models and an absorbed power law representative of :\par
\noindent
i) a binary disk population ($kT=5$~keV, metal abundances 0.4 
times solar).\par
\noindent
ii) a very soft diffuse thermal 
($kT=0.29^{+0.11}_{-0.07}$ ~keV, metal abundance 0.4 time solar);\par
\noindent
iii) the jet emission ($\Gamma=2.3$).
A recent study of extranuclear components of the jet with Chandra
(Kraft et al. 2002) has resolved the jet in several knots
 well fitted by absorbed power-laws (\nh=$1-7\times10^{21}$ cm$^{-2}$) 
with photon indices $\sim$2.0-2.5, in agreement with ROSAT.

Given the spectral/spatial resolution of \sax, we considered 
the Rosat/ASCA parameterization sufficiently accurate
for our analysis  and focused our attention on the nuclear component.
We used the two thermal components and the jet power law
for the soft spectrum (only the normalizations of each component could 
vary) and fitted the continuum with an absorbed power law.
The fit was initially unsatisfactory
($\chi^2=180$ for 117 d.o.f.), but we obtained an immediate improvement 
($\chi^2=142$ for 117 d.o.f.) when the temperature of 
the softer component and the column density of the jet 
were left free to vary. (The fit was insensitive to the 
temperature of the binary emission which was held fixed).
While the soft part of the spectrum 
appeared now adequately fitted, significant deviations of the residuals 
remained in the hard part of the spectrum. 
An excess of emission above 8 keV and a deficit above 80 keV were 
clearly visible in the MECS and PDS data.
We interpreted that as an indication of a possible bending of the 
spectrum and multiplied the power law by an exponential cut-off. 
The $\chi^2$ became immediately acceptable ($\chi^2$=125.6 for 116 d.o.f),
corresponding to an improvement of $99.985\%$. 
We also looked for a reflection
component, which might be due to an optically thick slab of gas
and be therefore associated with the iron line.
We used the PEXRAV model in XSPEC and fixed the inclination angle of the slab
to 65 degrees, an average value deduced from current estimates 
of the inclination of the jet (Jones et al. 1996; Israel 1998; Tingay 1998). 
The relative amount of reflection, Ref, turned out to be negligible, namely $<0.1$.
The fit results are reported in Table 3.

From a purely statistical point of view, another model,
namely two power laws absorbed by different values of \nh 
reproduced equally well the spectral curvature in the MECS and PDS.
A similar approach was followed by Turner et al (1997) to fit the 
ASCA data. However, we regard the previous model preferable,
not only for its similarity to the most commonly used description
in Seyfert galaxies, but also because, as shown in Section 6, in the
quasi-simultaneous observations of \sax and COMPTEL combined, 
the presence of a high energy break appears unequivocal. 

Finally, we briefly comment our results on the extended 
soft emission. Comparing our fits of the extended soft components with the
results of ROSAT (Turner et al. 1997) and Chandra (Kraft 2002) 
it appears that \sax data require a slightly higher temperature of the 
diffuse  component (ii) and a larger intrinsic \nh (and flux) of the jet (iii).
We think that this simply reflects the \sax  spatial limitations in 
resolving the several components present within a region of 4'. 
It is possible that we are including more inner (subparsec) jet knots 
or we are neglecting a un-resolved soft component .
For example,  if another  thermal component  (kT$\sim 0.7$ keV, 
F$_{0.4-2.0~keV} \sim 7\times10^{-12}$ erg cm$^{-2}$ sec$^{-1}$)
absorbed by Galactic \nh is added to the \sax fit, the nuclear parameters do
not change, but both the temperature of the extended component
and the \nh  of the jet are reduced, 
making  the agreement with the previous measurements better. 
However the $\chi^2$ improvement is 
modest ($\Delta\chi^2=6$ for three new parameters) and the F-test 
probability less than $95\%$.
The new soft component is not statistically necessary.
Although frustrating, this outcome is however useful.
It shows that somewhat different parameterizations of the soft emission do not 
impact on the results concerning the hard nuclear component ($>4$ keV). 
This assures that the simple parameterization of 
the soft spectrum (i.e. two thermal models and an absorbed power law),
adopted in the following, does not affect our conclusions on the high energy 
part of the spectrum.

\begin{deluxetable}{lccc}
\tabletypesize{\footnotesize}
\tablewidth{0pc}
\tablecolumns{4}
\tablecaption{Average Spectrum}
\tablehead{\multicolumn{4}{c}{Fits  to 0.4-250~keV combined \sax data}
}
\startdata
\multicolumn{4}{c}{\bf Extended  Emission}\\
&&&\\
N$_H$  [$10^{21} cm^{-2}$]  &0.7(f) \\
kT[keV]                     &0.61$^{+0.08}_{-0.12}$  \\
Flux [0.4-2.0 keV]\tablenotemark{a} &0.14$^{+0.06}_{-0.03}$\\
&&&\\
N$_H$  [$10^{21} cm^{-2}$]  &0.7(f) \\
kT[keV]                     &5.0(f) \\
Flux [0.4-2.0 keV]\tablenotemark{a} &0.14$^{+0.08}_{-0.11}$\\
&&&\\
N$_H^{\it Jet}$  [$10^{22} cm^{-2}$]   &2.3$^{+3.8}_{-1.2}$\\
$\Gamma^{\it Jet}$           &2.3(f) \\ 
Flux$^{\it Jet}$  [0.4-2 keV]\tablenotemark{a}         & 1.0$^{+5.6}_{-0.6}$\\
Flux$^{\it Jet}$  [2-10  keV]\tablenotemark{a}         & 0.7$^{+4.1}_{-0.4}$ \\
&&&\\
\multicolumn{4}{c}{\bf Nuclear Emission}\\
&&&\\
$N_H^{\it Nucl.}$  [$10^{21} cm^{-2}$] &102$^{+9}_{-4}$ \\
$\Gamma^{\it Nucl.}$                    &1.80$^{+0.03}_{-0.04}$\\
E$_{\rm Cutoff}$ [keV]     &621$^{+677}_{-214}$\\
Ref. &$<0.10$  \\
Flux$^{\it Nucl.}$ [2-10 keV]\tablenotemark{a}      & 38$^{+3}_{-4}$\\
Flux$^{\it Nucl.}$ [15-250 keV]\tablenotemark{a}    & 114$^{+7}_{-14}$\\
&&&\\
$\chi^2$(dof)$^b$&126 (114) \\
\hline
&&&\\
\multicolumn{4}{c}{\bf Iron Line}\\

&&&\\
&\multicolumn{2}{c}{Cold Lines}& Ionized Line \\
& $K_{\alpha}$&$K_{\beta}$&\\
&&&\\
E [keV]    &6.4 (f)& 7.1 (f)& 6.8$^{+0.4}_{-0.3}$ \\ 
$\sigma$  [keV]       &$0<0.13$& $0<0.13$ & 0.31$^{+0.30}_{-0.31}$ \\ 
I\tablenotemark{c}  &2.8$^{+0.8}_{-1.4}$&0.3$^{+0.1}_{-0.1}$& 1.4$^{+1.3}_{-0.8}$ \\
EW (eV)     &69$^{+21}_{-34}$ &9$^{+3}_{-5}$ &39$^{+36}_{-22}$  \\
&&&\\
$\chi^2$(dof)               &150(142)&&\\
\enddata
\tablenotetext{a}{Unabsorbed flux ($\times 10^{-11}$ erg cm$^{-2}$ sec$^{-1}$).}
\tablenotetext{b}{$\chi^2$ values refer to the fits performed excluding
the 5.5-7.5 keV iron line region}
\tablenotetext{c}{Intensity of the iron line in units of $10^{-4}$ 
photons cm$^{-2}$ sec$^{-1}$.}
\end{deluxetable}

\subsection{\it The Iron Line Feature}

Once the shape of the hard continuum is established, we could study
the iron line feature\footnote{With the term ''feature'' we describe
all the Fe photons in the 5.5-7.5 keV band. When we refer to one of the 
component of the feature, we use the term ''line''}.
The residuals in the 5.5-7.5 keV region (Figure 3) show 
a complex feature in emission  with an evident blue tail.
We note, incidentally, that this asymmetry of the line was also 
visible in the ASCA data (Figure 6 in Turner et al. 1997), though 
the hard component had been fitted with a very complex model consisting of 
three differently absorbed power laws.
\begin{figure}
\begin{center}
\vbox{\psfig{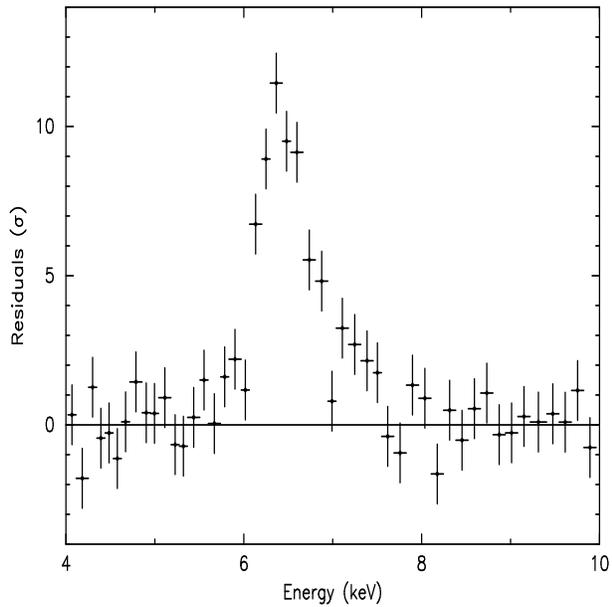}
}
\vspace{-0.1cm}
\caption{\footnotesize{Residuals in the in iron 
line region of the \cena average spectrum. The underling nuclear 
continuum is fitted with an absorbed  cutoffed power law.
A blue tail is clearly present.
}}
\end{center}
\end{figure}

In order to better understand the structure of the feature, we fixed 
all the parameters of the soft components and the hard continuum and fitted 
the 5.5-7.5 keV excess with a simple gaussian profile. 
The  $\chi^2$ value ($\chi^2= 162$ for 144 d.o.f) and an inspection of the 
residuals showed again an excess around 7 keV.
A single cold ($E=6.50^{+0.05}_{-0.04}$ keV), moderately broad 
($\sigma=0.18^{+0.07}_{-0.09}$ keV) line was not sufficient to model 
properly the whole profile.
Then we tested whether the blue tail was produced by the  Fe $K_{\beta}$ 
fluorescence line and a  second gaussian was added to the fit.
The energies of the two 
lines were frozen at 6.4 keV and 7.1 keV and  
the intrinsic widths were forced to 
vary together ($\sigma_{K_{\alpha}}=\sigma_{K_{\beta}}$).
Finally the intensities were linked in order to 
reproduce the intensity ratio of 150 to 17. 
The $\chi^2$ became worse ($\chi^2= 169$ for 145 d.o.f) and the blue tail was 
still present in the residuals. 
Only the inclusion of a third gaussian, with all its parameters free, 
gave a satisfactory modeling of the 7 keV excess and a $\chi^2$ of 
150 for 142 d.o.f, with a significant 
improvement in comparison with the previous cases (P$_{F_{test}}>99.6\%$).
Although the $K_{\beta}$ line was not statistically necessary, we kept
its contribution linked to the $K_{\alpha}$ line, in order to more
correctly evaluate the intensity of the ionized line.

We conclude that, in addition to the $K_{\alpha}$ 
line, in our observations \cena shows an ionized line in emission.
In particular, as displayed by the  $E-\sigma$ confidence levels of the 
third (hot) component (Figure 4), the state of ionization 
of the Fe atoms responsible of the blue tail is greater than XX (i.e. 
$E>6.5$ keV) at the 90$\%$ confidence level.

\begin{figure}
\begin{center}
\vbox{\psfig{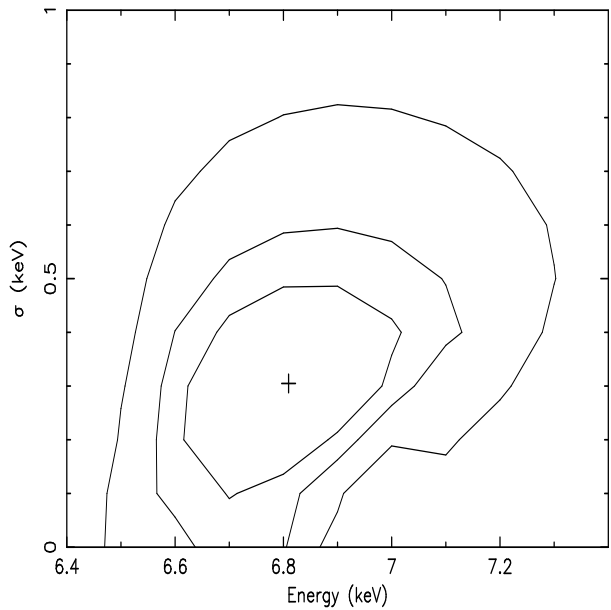}
}
\vspace{-0.1cm}
\caption{$E$-$\sigma$ contour plot (68$\%$, 90$\%$, 99$\%$ confidence levels)
for the ionized component  required by \sax data in addition to the cold 
K$_{\alpha}$ and  K$_{\beta}$ fluorescence lines
}
\end{center}
\end{figure}

The results of our analysis are summarized in Table 3.
The data, the residuals and the final model for the 0.4-250 keV 
spectrum are shown in Figure 5. 
\begin{figure}
\begin{center}
\vbox{\psfig{figure=figure5a.ps,height=8.0cm,width=8.0cm,angle=-90}
\psfig{figure=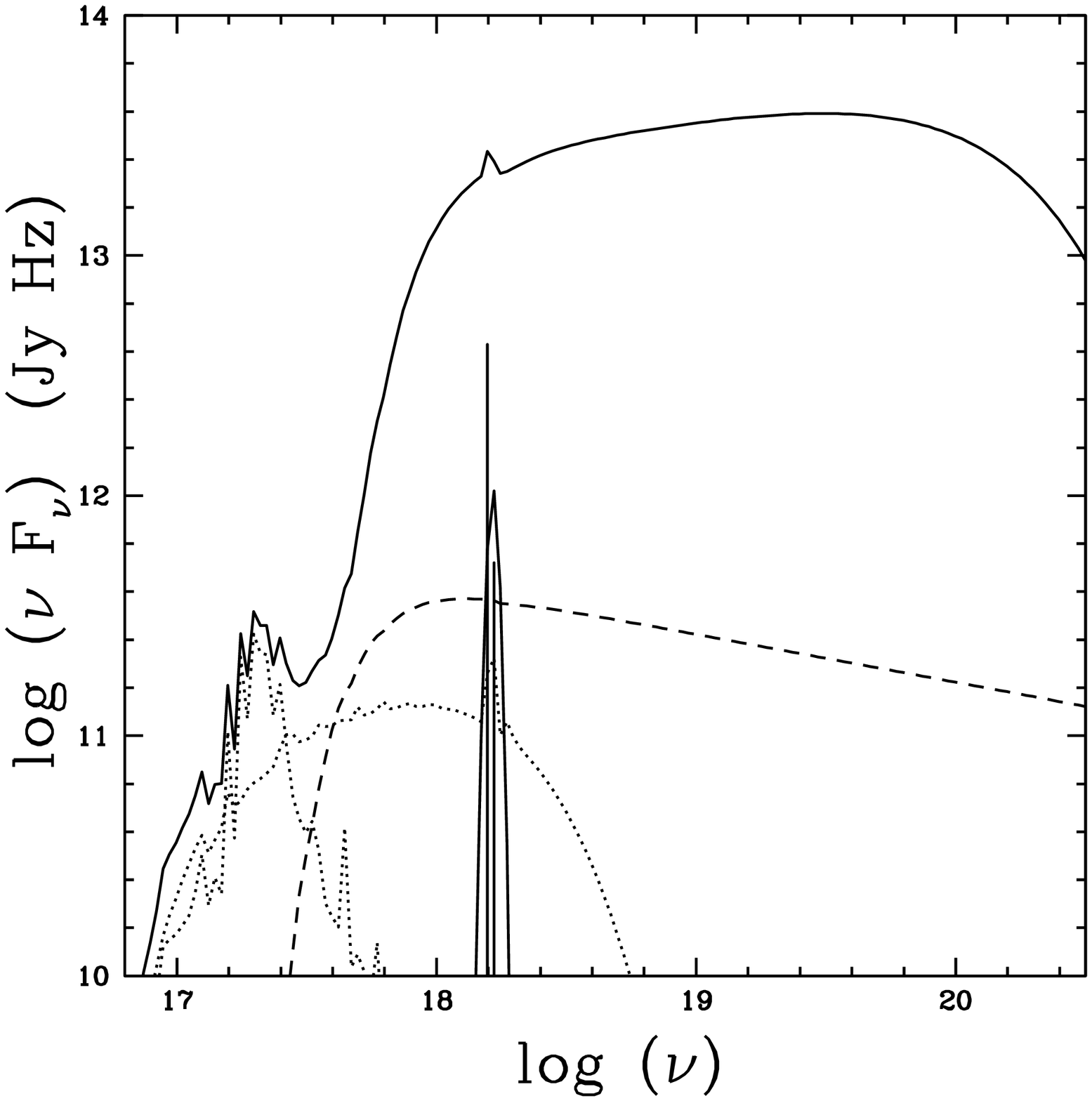,height=8.0cm,width=8.0cm}
}
\vspace{-0.1cm}
\caption{\footnotesize{{\it Top Figure} -- 
The average spectrum and the residuals of all the 5 
\sax observations combined together.
({\it Bottom Figure}) -- The simplest model used to fit the data.
It consists of two extended thermal emissions (dotted lines), 
a jet emission (dashed line), a strongly absorbed nuclear power 
law with high energy cutoff (solid line) and  three gaussian profiles
corresponding to the $K_{\alpha}$ and $K_{\beta}$ cold features plus a 
ionized component. 
}}
\end{center}
\end{figure}

\subsection{\it \sax and COMPTEL Simultaneous Observations}

The overlapping of the \sax and \comptel pointings during the summer 1999
offers the unusual opportunity to study the shape of the continuum 
of emission on more than 4 decades in energies from 0.4 keV to 30 MeV.
In Figure 6 the COMPTEL data are plotted together with 
the two \sax spectra collected in the summer of 1999.
The presence of a high energy break in the \cena spectrum is unequivocal.
Incidentally, we note that our results are in agreement with Kinzer et al. 1995,
which observed a cutoff at about 500-900 keV with OSSE/GRO 
when the source was in a state of brightness 
similar to that of \sax (F$_{100 keV}\sim 2.2$ cm$^{-2}$ sec$^{-1}$ MeV$^{-1}$).

\begin{figure}
\begin{center}
\vbox{
\psfig{figure=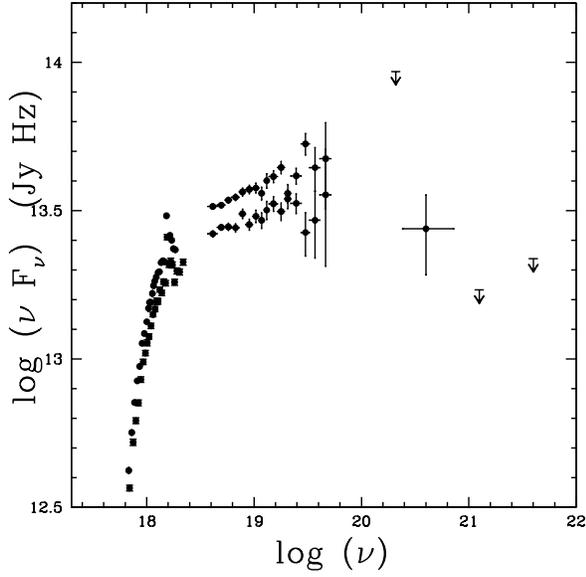,height=8.0cm,width=8.0cm}
}
\vspace{-0.1cm}
\caption{\footnotesize{The \sax and COMPTEL quasi-simultaneous observations
performed in the summer of 1999. 
}}
\end{center}
\end{figure}

\section{Spectral Variability}

\subsection{\it The Nuclear Continuum}

We initially looked for long-term
spectral variability using the softness (SR:1.5-4.0/4.0-10 KeV) 
and the hardness (HR:15-250/4-10 keV) ratios.
As for to the time variability study, a spectral change was accepted when 
$P_{\chi^2}< 10^{-3}$.
As shown in Figure 8, the SR increased 
with the nuclear intensity ($P_{\chi^2_{SR}}\sim 10^{-6}$), indicating a 
clear spectral modification below 4 keV.
On the contrary, the HR did not  meet the variability criteria 
($P_{\chi^2_{HR}}=0.008$), implying
no significant spectral changes of the continuum above 4 keV.
\begin{figure}
\begin{center}
\vbox{\psfig{figure=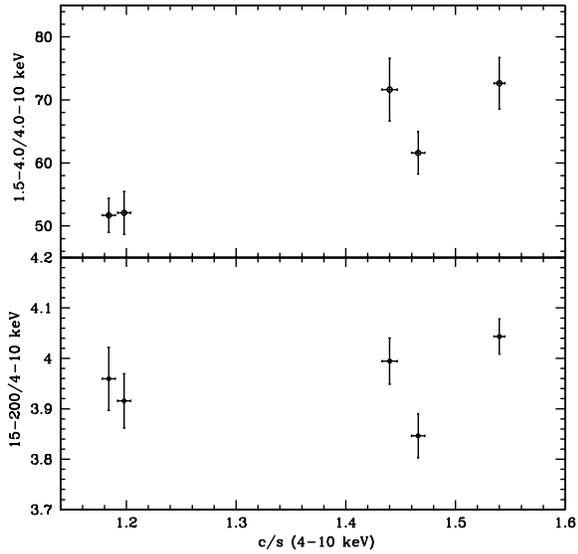,height=8.0cm,width=8.0cm}
}
\vspace{-0.1cm}
\caption{\footnotesize{{\it Upper Panel} -- 
Softness Ratio (1.5-4 keV/4.10 keV) versus 4-10 keV count rates.
({\it Lower Panel}) -- Hardness Ratio (15-250 kev/4-10 keV) versus 
4-10 keV count rates}}
\end{center}
\end{figure}

In order to understand the nature of the observed spectral variability,  
we initially fitted each \sax observation individually, as we had done for
their sum, excluding the region of the iron line. 
The large number of free parameters, though,  prevented us from detecting 
significant spectral modifications by directly comparing the fit results.
Then we decided to adopt a different  strategy.
We fitted the five observations simultaneously, with all the parameters
linked except for the normalization of the 
cut-offed power law ($\chi^2=$ 689 for 652 
degrees of freedom). As the next step, we left the parameters \nh, $\Gamma$ and $E_{cutoff}$, 
one at a time, free to vary from one observation to the other.
If, for a given variable,  the fit improvement was better than 
$99.5\%$, we concluded that 
the relaxed parameter had  changed during the \sax monitoring.

In agreement with the HR study, 
neither the spectral slope ($\chi^2=$ 672 for 647 d.o.f.) 
nor the high energy cutoff ($\chi^2=$ 679 for 647 d.o.f.) significantly 
varied from 1997 to 2000 in spite of the continuum flux variations.
Conversely, a substantial drop in $\chi^2$ (see Table 4) was 
obtained as a consequence of, albeit modest, modifications in
the column density. In particular the gas appeared thinner when 
the nucleus luminosity increased (compare the 1998 results with the 
other epochs).
Note that \nh modifications have been also found by  Benlloch et al. (2001) 
during a RXTE monitoring of \cena performed from 1996 to  2000.
However, although small changes in \nh could easily explain the observed Softness Ratio 
variability (Figure 7), we do not regard this conclusion to be very robust.
Flux  variations of unresolved soft 
components could actually mimic changes of the  column density 
along the direct path. For example, if the  jet normalization 
was let free to vary (and \nh kept frozen),  
the fit improvement was, also in this case, 
larger than 99.5$\%$ ($\chi^2=$ 636 for 647 d.o.f.).

\begin{deluxetable}{lccccc}
\tabletypesize{\footnotesize}
\tablewidth{0pc}
\tablecolumns{6}
\tablecaption{Variability Study Results}
\startdata
  & 1997& 1998& July 1999 & August 1999& 2000\\
&&&&&\\
\multicolumn{6}{c}{\bf Nucleus}\\
&&&&&\\
\multicolumn{1}{l}{\nh [10$^{21}$ cm$^{-2}$]}&
102$^{+2}_{-2}$&
98$^{+1}_{-2}$ &
100$^{+1}_{-2}$&
101$^{+2}_{-2}$&
100$^{+2}_{-1}$\\
$\Gamma$ 
& 1.80(f) &1.80(f) & 1.80(f) &1.80(f) &1.80(f) \\
$E_{Cutoff}$ [keV] & 621(f) & 621(f) &621(f) &621(f) 
&621(f)\\
\multicolumn{1}{l}{Flux [2-10 keV]$^a$}&
32.1$^{+0.5}_{-0.5}$&
42.9$^{+0.4}_{-0.6}$&
39.8$^{+0.6}_{-0.7}$&
32.2$^{+0.7}_{-0.5}$&
39.3$^{+0.6}_{-0.7}$\\
&&&&&\\
\multicolumn{1}{l}{$\chi^2$ (d.o.f)}&\multicolumn{3}{l}{639 (647)}\\
\hline
&&&&&\\
&&&&&\\
\multicolumn{6}{c}{\bf Iron Feature}\\
&&&&&\\
E [keV]&
6.4(f)&
6.4(f)&
6.4(f)&
6.4(f)&
6.4(f)\\
$\sigma$  [keV]   &
0 (f) &
0 (f) &
0 (f)&
0 (f) &
0 (f)\\
\multicolumn{1}{l}{I$^b$}&
2.8$^{+0.9}_{-0.8}$&
2.1$^{+0.8}_{-0.7}$&
3.1$^{+1.1}_{-1.2}$&
2.1$^{+1.1}_{-1.0}$&
3.9$^{+1.2}_{-1.2}$\\
&&&&&\\
E [keV]&
6.8(f)&
6.8(f)&
6.8(f)&
6.8(f)&
6.8(f)\\
$\sigma$  [keV]   &
0.3 (f) &
0.3 (f) &
0.3 (f)&
0.3 (f) &
0.3 (f)\\
\multicolumn{1}{l}{I$^b$}&
1.7$^{+1.0}_{-1.0}$&
0$<1.0$&
2.7$^{+1.1}_{-1.3}$&
2.7$^{+1.1}_{-1.2}$&
0.8$^{+1.4}_{-0.8}$\\
&&&&&\\
\multicolumn{1}{l}{$\chi^2$ (d.o.f)}&
\multicolumn{3}{l}{736 (750)}\\
\enddata
\tablenotetext{a}{Unabsorbed flux ($\times 10^{-11}$ erg cm$^{-2}$ sec$^{-1}$).}
\tablenotetext{b}{Intensity of the iron line in units of $10^{-4}$ photons cm$^{-2}$ sec$^{-1}$.}

\end{deluxetable}

\subsection{\it Iron Line Variability}

The iron line feature changed intensity during the \sax monitoring.
This was immediately clear at the beginning of the monitoring campaign, 
when the first two observations were compared.
Figure~8 shows the ratio between the 1997 and 1998 spectra.
It is evident that the ratio is constant across the entire
3-150~keV range apart from the iron line region. The feature was  
clearly more intense when the source was fainter (in 1997).

We therefore attempted to investigate the behaviour of the cold
and of the ionized lines separately, following the same procedure adopted  
for the nuclear variability. 
A simultaneous fit of the 5 \sax observations was performed including 
also the iron line region. All the parameters of 
the extended components and the nuclear emission were frozen.
To take into account the continuum variability, we choose to fix 
for each observation the correspondent column density and nuclear 
flux reported in table 4.
\begin{figure}[h]
\begin{center}
\vbox{\psfig{figure=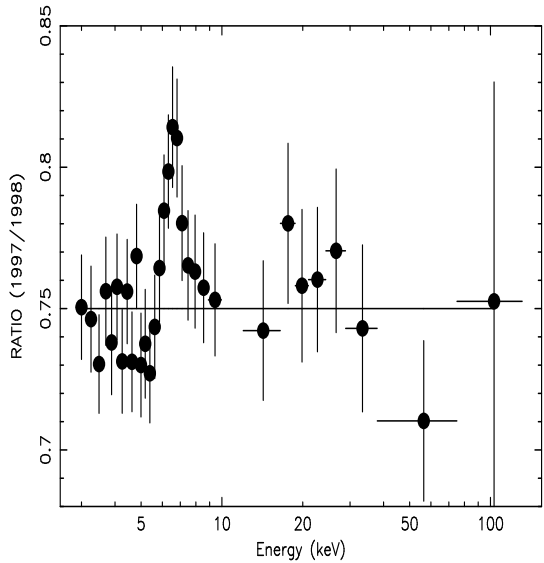,height=8.0cm,width=8.0cm,angle=-90}
}
\vspace{-0.1cm}
\caption{\footnotesize{ The 1997 data divided by the 1998
data. The ratio is constant across the 3-150~keV band
but not in the 5-7~keV iron-line range.
}}
\end{center}
\end{figure}

At first, the intensities of the three gaussian were kept frozen to 
their average values reported in Table 3.
The obtained $\chi^2$ is 783 for 760 degrees of freedom.
The we relaxed the normalization of the 6.4 keV (to which the $K_{\beta}$
is linked) and the 6.8 keV gaussians. The fit improved with a significance
of more than 99.9$\%$ (see Table 4), thus confirming the time 
variability of the feature. These results show that,
while the cold line did not significantly change intensity over the 
whole campaign, the ionized line appears to have been the one responsible 
for the variations in the feature. In particular, the ionized line
was not detected in 1998, when the X-ray source was in the brightest state.
Moreover, when the two 1999
observations, performed 23 days apart, are compared, we note that the
measured intensity is consistent with a constant value (in fact the largest
among the 5 observations), while the continuum strength in the first was
only slightly less than in the 1998 observation, and dropped by $\sim 20\%$ 
in the second. 
Thus we conclude that the variations in the ionized line do
not seem to correlate with the nuclear luminosity.

\section{Discussion}

The study of \cena is clearly complex. In particular, the 
soft part of the spectrum ($<4$ keV) is rich in components that we could 
not directly resolve with the imaging capability of \sax.
Our spectral analysis was therefore essentially focused on the study of the 
medium and hard X-ray spectrum, although the main extranuclear extending
components, namely the galactic ridge, the extended soft thermal emission 
and the kpc jet, were included in our fits.

The results of this paper can be summarized in four points:\\ 
i) definition of the hard continuum, i.e. 
determination of a sharp curvature at $\sim1$ MeV and  
of a stringent upper limit consistent with the absence of a cold 
reflection component; \\
\noindent 
ii) discovery of an ionized, in addition to a cold $K_{\alpha}$
iron line;\\
\noindent
iii) detection of iron line variability, due mainly to changes in
the ionized component, which do not appear to correlate with those
in the strength of the nuclear continuum;\\
\noindent
iv) detection of spectral variations below 4 keV.\\

The nature of the nuclear 
emission in \cena is still unclear and strongly 
debated. Chiaberge Capetti $\&$ Celotti (2001) noted that the 
Spectral Energy Distribution (SED) of the nucleus is very similar to that  
of Blazars, showing two broad peaks, located in the far-infrared band 
and at about 100 keV.  
They fitted the \cena SED with a  Synchrotron Self Compton Model and found 
a beaming factor $\delta=[\Gamma(1-\beta cos\theta)]^{-1}\sim 1$ which, with 
$cos\theta\sim0.4$, turned out to be 
incompatible with being originated in a misoriented jet 
(assuming bulk Lorentz factors $\Gamma\sim 15-20$, as estimated in BL Lacs).
The idea suggested by the authors is that there is a distribution in the bulk
velocity of the jet with a slower external layer surrounding a highly relativistic 
inner flow. The SED of \cena (and more in general of FRI galaxies, 
Chiaberge et al. 2000)  should be dominated by the slower plasma emission.
The Fe lines observed by \sax
could be produced by interaction between external jet X-ray photons
and surrounding neutral/ionized gas.
A very wide opening angle for the non-thermal plasma ejected by 
the nucleus is then required to explain the prominent feature in emission.

An alternative interpretation of the SED 
is that the main source of medium-hard X-ray photons is an 
(isotropically emitting) accretion flow.
The high energy spectral 
curvature observed by \sax and COMPTEL  could be simply due to 
up-scattering of soft photons by thermal 
($E_{cutoff}\propto kT$) electrons in an accretion flow.
Incidentally, we note that Verdoes Kleijn et al. (2002), on the basis of nuclear 
radio and optical  observations of a sample of 21 FRI nearby radio galaxies, 
have drawn  similar conclusions.
In order to explain the tight correlation between radio, optical
and (isotropic) H$alpha$+[N II] core emission, they suggested
the possible presence of an accretion flow in addition to a jet.

If an accretion flow dominates the X-ray continuum of \cena, 
it is not probably in the shape of the 
standard cold thin optically thick 
disk (SS; Shakura \& Sunyaev 1973) with a hot corona above it 
(Haardt $\&$ Maraschi 1991, 1993; Poutanen $\&$ Swenson 1996).
The lack of a significant reflection component as well as the presence of a 
narrow $K_{\alpha}$ line weakens this possibility.
As inferred from the 6.4 keV gaussian profile, the neutral gas is 
tens of light-days distant from the black hole.
Assuming a black hole mass of $M=2\times 10^{8} M_{\odot}$
(Marconi et al. 2001), the upper limit $\sigma_{6.4~keV}=0.13$ keV of the 
$K_{\alpha}$ width (Table 3) implies indeed a distance of the neutral zone 
larger than $R_{6.4~keV}>1.4 \times 10^{17}$ cm ($>0.05$ pc).
The large uncertainties associated with the 6.8 keV gaussian parameters prevent
us from excluding that an ionized disk extends up to a few Schwarzschild radii
(R$_s=2GM/c^2\sim 6\times10^{13}$) from the black hole. 
Considering 
the low accretion rate of \cena $\dot m \leq 10^{-3} \dot m_{\rm Edd}$
(as deduced by the SED reported by Chiaberge et al. 2001, 
assuming a central  mass of $M=2\times 10^{8} M_{\odot}$), 
this is however improbable. 
The ionization of the upper layers of a SS disk strongly depends on the
accretion rate (Ross and Fabian 1993). As shown by the accurate models of 
Nayakshin and Kallman (2001), for such a modest value of  $\dot m$ 
the disk should be neutral for a wide range of illumination 
conditions (i.e. corona geometries).
In addition,  the iron line variability does not support the idea that 
the reprocessing material is in proximity of the X-ray primary source.
The lack of correlation between the nuclear 
luminosity and the ionized line suggests a 
a temporal delay and therefore a spatial 
separation between the X-ray source and the gas. 

Incidentally we note that the 6.8 keV variability could be 
explained without invoking a temporal delay.
If the contributions of the nuclear thermal and 
non-thermal radiation are comparable, then   
the variability of the X-ray continuum 
could be the result of random and un-related changes 
of the jet and the accretion flow. 
For example, a very faint accretion flow (responsible for the Fe line
production) and a very bright jet could easily account for the weaker 
Fe feature and the intense X-ray continuum observed in 1998.
Although interesting, this interpretation is not convincing.
It is difficult to justify the lack of significant spectral modifications
(slope and energy cutoff) of the continuum on time scale of years
if intense and sometimes opposite changes of the luminosity of its components
occurred from one observation to another.

In any case, the observation of a iron feature which does not promptly 
respond to the X-ray source variations is not a surprising result.
Studies of X-ray bright radio galaxies with \sax and RXTE have
shown a general weakness of the iron line and the
reflection (Eracleous et al. 2000, 
Zdziarski and Grandi 2001, Grandi et al. 2001).
One possible explanation is that the covering factor of the 
reprocessing gas is small, since  the SS cold disk switches to a 
hot thick optically thin accretion 
flow in its inner region (Shapiro Lightman and Eardley (1976),
Narayan $\&$ Yi 1995).  
If this is the case, a temporal delay between 
line (produced from the disk) and continuum (emitted form the hot flow)
is expected.
In \cena, characterized by a low radiative
efficiency, the inner regions could be occupied by an 
Advection Dominated Accretion Flow, i.e. very hot thin plasma which, unable to 
cool efficiently, advects most of its thermal 
energy into the central black hole.

Finally, we note that the spectral variability study (Table 4) 
indicates a larger column density when the Fe feature was intense (and the nucleus less luminous). 
If the same gas which absorbs the nuclear continuum also produces the lines, then the
drastic reduction of Fe photons when the nucleus was brighter could imply a larger transparency of 
the medium.
Although intriguing, a correlation between \nh and Fe intensity changes is unlikely 
for several reasons.
The variations ($15-20\%$) of the nucleus luminosity are too modest to signifcatively 
alter the state of ionization of the gas. 
In addition, Benlloch et al. 2001, on the base of RXTE data, 
showed that the \nh variations, observed in three distinct pointings, 
are correlated neither with the continuum flux nor with the iron line.
Finally, the column density decrease could be not real, but mimicked by the flux increase of 
unknown (and unresolved) soft components, like, 
for example, jet knots, variable stars within the \sax field 
and/or radiation scattered into our line of sight.

\end{document}